\documentstyle[aps,preprint,prc]{revtex}
\date{\today}
\begin{document}
\flushbottom


\widetext
\draft
\title{Strange Neutral Currents in Nuclei}
\author{M.T. Ressell}
\address{
W.K. Kellogg Radiation Laboratory, 106-38,
California Institute of Technology,
Pasadena, CA 91125}
\author{G.J. Mathews}
\address{
Department of Physics,
University of Notre Dame,
Notre Dame, IN 46556}
\author{M.B. Aufderheide}
\address{
Department of Physics,
University of Pennsylvannia,
Philadelphia, PA 19104 \protect \\
and University of California,
Lawrence Livermore National Laboratory,
Livermore, CA 94550}
\author{S.D. Bloom and D.A. Resler}
\address{University of California,
Lawrence Livermore National Laboratory,
Livermore, CA 94550}

\date{\today}
\maketitle

\def\thepage{\arabic{page}}
\makeatletter
\global\@specialpagefalse
\ifnum\c@page=1
\def\@oddhead{Draft\hfill To be submitted to Phys. Rev. C}
\else
\def\@oddhead{\hfill}
\fi
\let\@evenhead\@oddhead
\def\@oddfoot{\reset@font\rm\hfill \thepage \hfill}
\let\@evenfoot\@oddfoot
\makeatother

\begin{abstract} We examine the effects on the nuclear neutral
current Gamow-Teller (GT) strength of a finite contribution from
a polarized strange quark sea.
We perform nuclear shell
model calculations of the neutral current GT strength for a number
of nuclei likely to be present during stellar core collapse.  We compare
the GT strength when a finite strange quark contribution is included
to the strength without such a contribution.  As an example, the
process of neutral current nuclear de-excitation via
$\nu {\overline{\nu}}$ pair production is examined for the two cases.
\end{abstract}

\makeatletter
\global\@specialpagefalse
\def\@oddhead{\hfill}
\let\@evenhead\@oddhead
\makeatother
\nopagebreak
\narrowtext
\section{INTRODUCTION} \label{sec:intro}

A number of recent experiments have provided tantalizing hints
that the strange quark sea within the nucleon may play a major role in
determining its physical properties.  Most notably, the strange quarks may
be polarized and a major contributor to the spin of
the nucleon \cite{emc,smc}.
(The fraction of the proton's spin carried by the strange quark sea
is usually denoted by $\Delta s$.)
This interpretation is somewhat controversial
\cite{jaffe},
but remains the favored explanation of the experiments which
measure the spin distribution of the nucleon.  If the
strange quark sea is polarized and contributes significantly
to the nucleon's spin, then there are numerous implications for
particle and nuclear physics as well as astrophysics.
Among these are effects on neutral current interactions
\cite{kaplan,beise,behr,sky,garvey} and
more exotic effects on processes such as neutralino-nucleus
scattering (which is of fundamental importance in the search
for particle dark matter \cite{ress93}).  In this paper, we
focus upon inelastic neutrino-Nucleus, $\nu A$, interactions.  In
particular, we examine the effects of $\Delta s \ne 0$
in the nucleon upon the neutral current Gamow-Teller (GT) strength
in a number of nuclei that are present during supernova collapse.

The GT operator results from the axial-vector current in the
non-relativistic and zero momentum-transfer limits (i.e. the
allowed approximation).  It is
the dominant contribution in $\nu A$ inelastic scattering.  In the limit
where strange quarks do not contribute to nucleonic properties
(the standard case), the operator is purely isovector in nature.
The inclusion of nucleonic spin due to polarized strange quarks,
however, leads
to an important change  in the form of the GT operator. It acquires
an {\it isoscalar} component!  The zeroth order effect of this change is
to increase the neutral current interaction strength of protons relative
to that of neutrons \cite{kolbe}.
This shift results in the redistribution
of the neutral current GT strength function, $B(GT_0)$, for a nucleus.
The inclusion of this isoscalar piece in the GT operator circumvents the
usual isospin selection rule which forbids $T = 0 \rightarrow T = 0$
transitions.  This can open new transition channels and lead to a
significant rearrangement of the low lying GT strength for $T = 0$
nuclei.  These new channels may present a method for a precise
measurement of $\Delta s$ \cite{sky,garvey,hkpw}.  We briefly discuss
this possibility in Section III.  Also of importance is the
effect of the redistribution of $B(GT_0)$ upon the neutral current
interaction rates. These rates are of interest as they may
affect the neutrino distribution in core
collapse supernovae.

It has been realized for some time that inelastic $\nu A$ processes
may play an important role in the pre- and post-collapse phases of
supernovae \cite{rocky,fm,bh}.  The interaction rates for all of these
inelastic processes are highly energy dependent and hence quite
sensitive to the exact distribution of GT strength in nuclei.  The
neutral current GT strength distribution undergoes significant shifts
if $\Delta s$ differs appreciably from zero.  Hence, the effects of
strange quarks in the nucleon could have profound effects upon
supernova dynamics.  The effects of a non-zero value of $\Delta s$
upon $\nu$-process nucleosynthesis \cite{woos} has previously been
studied for several nuclei in the continuum random phase approximation
\cite{kolbe}.  Here we examine allowed neutral current GT processes
which might play an important role in the heating and cooling of the
collapsing star.  As an example,
we concentrate upon the process of nuclear de-excitation via the
emission of a $\nu\overline\nu$ pair (the neutral current
analog of $\beta$-decay), $A^* \longrightarrow A \nu \overline\nu$.
Because of phase space considerations, this process should be especially
sensitive to $\Delta s$ as well as being straightforward to
calculate.

Supernova cores present an environment where nuclei may develop
a large neutron excess.  Since a finite $\Delta s$ increases the
strengths of $\nu p$, relative to $\nu n$, interactions there
could conceivably be significant changes in the GT strength
for nuclei with large neutron-proton asymmetry.
An effect which will tend to compensate for
this is Pauli blocking as the available neutron orbitals
are filled.  To investigate these two competing
effects we examine the changes in $B(GT_0)$ induced by a non-zero
value of $\Delta s$ in a series of iron isotopes with increasing
$N - Z$.  We will show that fairly significant changes in
$B(GT_0)$ can occur for very neutron rich nuclei.

Our calculations of $B(GT_0)$
are performed in the nuclear shell model.
This approach allows an accurate representation of the low
lying strength, which is of paramount importance at the temperatures
of interest.

\section{FORMALISM AND WAVE FUNCTIONS} \label{sec:formandwave}

The strange quark content of the nucleon plays a role in both the
vector and axial-vector pieces of the weak hadronic current.  The
full formalism is presented in refs. \cite{beise,garvey,kolbe}.  Here
we are concerned only with the axial-vector piece, which
has the form
\begin{equation}
J^A_{\mu} = G_1(Q^2) \gamma_{\mu} \gamma_5, \label{CURRENT}
\end{equation}
where
\begin{equation}
G_1(Q^2) = -{1\over{2}} G_A^3(Q^2) \tau_3 + {1\over{4}} G_A^s
(Q^2). \label{FF}
\end{equation}
$G_A^3$ is the usual isovector coupling constant $G_A^3(0) =
g_A = 1.262$ and $G_A^s$ is the isoscalar coupling arising because
of the strange quarks in the nucleon.  The EMC data implies a value
of $G_A^s(0) = -0.38 \pm 0.12$ \cite{emc,beise,kolbe} and we adopt
this value here.
Recent measurements by the SMC experiment
\cite{smc} have found a slightly lower value for $\Delta s$ and hence
$G_A^s$.  We use the larger, EMC value, to examine the maximum
effect of the strangeness in the nucleon upon $B(GT_0)$.
To make this even more pronounced, we quench the value of the isovector
piece by the canonical amount, setting $g_A = 1.0$.
The isospin operator has the values
$\tau_3 = +1 (-1)$ for protons (neutrons).  $G_1(Q^2)$ is assumed
to have the standard dipole form but this is irrelevant for this
work since we work in the $Q^2 \rightarrow 0$ limit.

In the zero momentum transfer and non-relativistic limits (valid
for most supernova neutrinos) eqs. (\ref{CURRENT},\ref{FF}) lead to
the neutral current GT operator
\begin{equation}
GT = -{1\over{2}} ( g_A \tau_3 - {1\over{2}} G_A^s ) \mbox{\boldmath $\sigma$}.
\label{GTop}\end{equation}
The familiar form is recovered if $G_A^s = 0$.  It is immediately
apparent that by including $G_A^s \ne 0$ the GT operator will shift the
relative strengths of the $\nu p$ and $\nu n$ interactions
which are mediated by it.  (e.g. taking
$g_A = 1$ and $G_A^s$ = -0.38,  we find $GT\vert_{protons} = -0.595
\mbox{\boldmath $\sigma$}$ and
$GT\vert_{neutrons} = 0.405 \mbox{\boldmath $\sigma$}$.)  Also note, that since
the operator is no longer purely isovector, $T = 0 \rightarrow
T = 0$ transitions are allowed and will be proportional to
$\vert G_A^s \vert^2$ \cite{sky,kolbe}.  Thus, we see that
the presence of a non-zero $G_A^s$ ~ can lead to potentially important
changes in the matrix elements for $\nu A$ interactions.
To determine if this is indeed the case, the modified GT operator,
eq. (\ref{GTop}), needs to be evaluated between realistic nuclear
wave functions and compared with the standard ($G_A^s$ = 0) results
for various nuclei.

To examine the effect of the modified GT operator, eq. (\ref{GTop}),
upon $\nu A$ reactions we have computed states for nuclei in the $p$, $sd$,
and $fp$ shells.  All of the wave functions and strength functions were
generated using the nuclear shell model code CRUNCHER \cite{resler}
and its auxiliary codes.  We have studied the $p$-shell nuclei
$^{12}$C and $^{14}$N using the Cohen-Kurath interaction \cite{cohen}.
The $sd$-shell nuclei $^{20}$Ne, $^{24}$Mg, and $^{28}$Si were all
investigated using the W, or universal $sd$, interaction
\cite{wildenthal}.  In these two shells full basis $0 \hbar \omega$
calculations were performed.

In the $fp$ shell, we performed calculations for $^{56}$Ni as well
as a large series of iron isotopes using the fpvh interaction
\cite{fpvh}.  This interaction has been shown to reproduce excited
state energy spectra \cite{fpvh} and charged current GT strength
functions \cite{aufderheide} with reasonable accuracy.  We looked at the
even-even isotopes of iron ranging from $^{50}$Fe to $^{66}$Fe
to study the effect of increasing $N - Z$ upon $B(GT_0)$.  Because of
the large dimensions of the $fp$-shell wave functions, we employed
truncated model spaces for all of these nuclei.  (A full basis
calculation of $^{56}$Fe would have a m-scheme dimension of
$\sim 5 \times 10^8$, far larger than can be accommodated using
conventional diagonalization techniques.)  In Table \ref{table1},
we present the
list of $fp$ shell nuclei examined in this work.  We also list
the model spaces considered and the m-scheme basis dimension for
both the parent and daughter nuclei.
Our daughter spaces are expanded so that we would satisfy the
standard charged current sum rule.

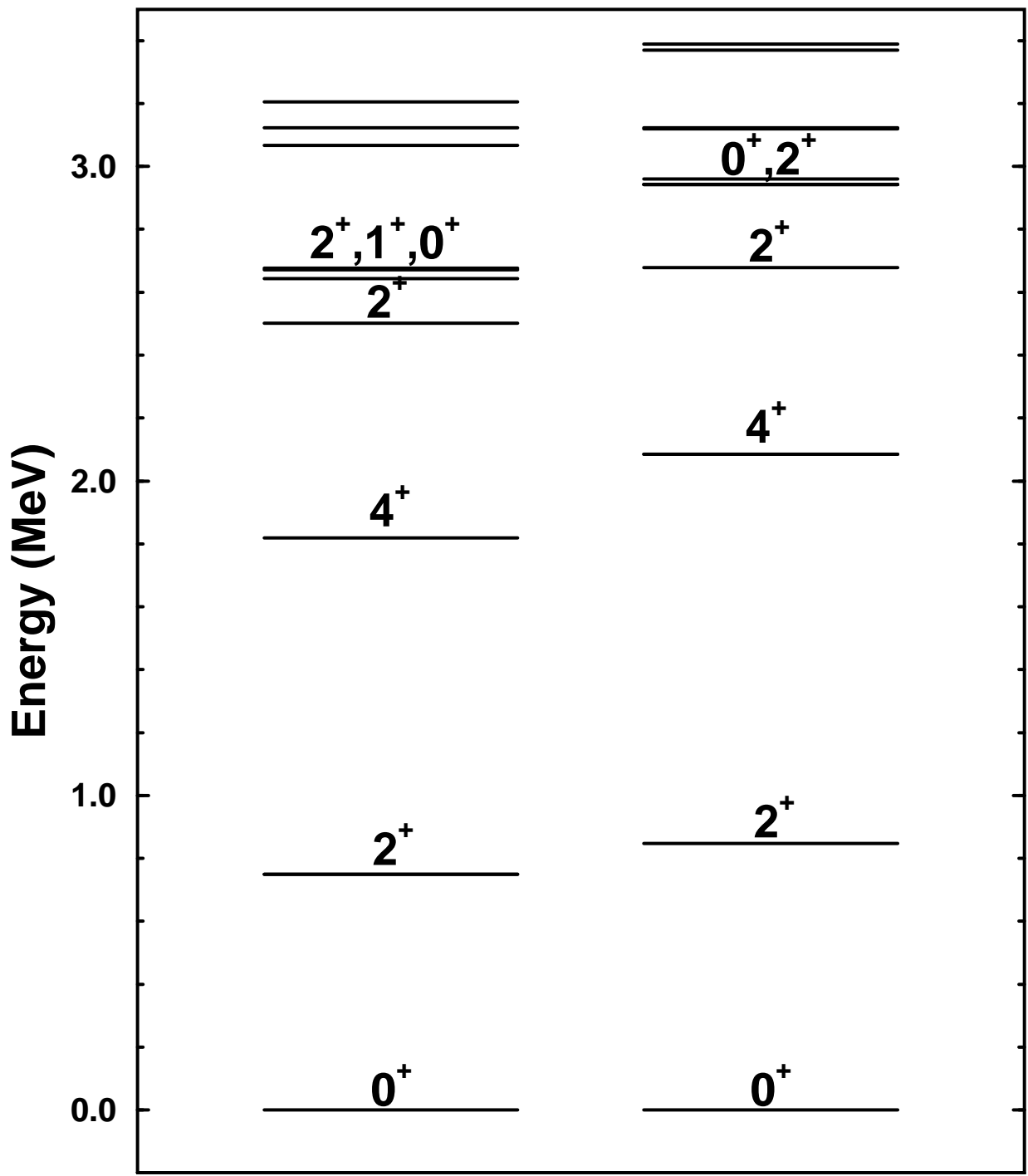
\begin{figure}
\caption[Figure 1]{The calculated (left) and measured \cite{lederer} (right)
excited state energy spectrum of $^{56}$Fe. The calculated spectrum
was obtained using the fpvh interaction \cite{fpvh} in the daughter model
space described in Table \ref{table1}.  The ground state has
$J^{\pi} = 0^+$.  The $J^{\pi}$ values of many of the low lying
states have been included for reference.}
\label{spectrum}
\end{figure}

In figure \ref{spectrum}, we present the calculated and experimental excited
state energy spectrum for the 10 lowest lying states of
$^{56}$Fe.  (The calculated spectrum is for the model space
with basis dimension 8738 in Table \ref{table1}.)  Figure 1 reveals
quite good agreement between theory and experiment and supports
the idea that the fpvh interaction in these model spaces produces
good wave functions for these nuclei.  Similarly good agreement
is obtained for other nuclei with measured spectra.

Another piece of evidence which lends credence to these
strength functions is the
good agreement between measured and calculated charged current
GT strength functions.  Aufderheide et al. \cite{aufderheide} have found
reasonable agreement between experiment and theory for several
$fp$-shell nuclei using the fpvh interaction in similar
model spaces.  The agreement is much better for $B(GT_-)$ than for
$B(GT_+)$ and, in any case,  is not perfect since it  requires
the usual quenching factor in order to match the magnitude
of the measured strength.  On the whole, the distribution of
strength is well reproduced in their calculations for sufficiently
large model spaces.  A similar
level of accuracy is expected to hold for the neutral current
processes calculated here.

\section{RESULTS} \label{sec:results}

Kolbe,
et al. \cite{kolbe} pointed out, that to first order, the ratio of the
proton to neutron cross section varies as
$\sigma_p/\sigma_n \approx 1 + 2 \vert G_A^s\vert/g_A$.
For $G_A^s = -0.38$ and $g_A = 1$ we find
$\sigma_p/\sigma_n = 1.76$ (for $g_A = 1.262$,
$\sigma_p/\sigma_n = 1.60$).  We see already that this effect
may be important.

The total GT strength for a given nucleus is
thought to scale roughly as \cite{fm}
\begin{equation}
B(GT_0) \propto \sum_{p,n} \sum_{i,f}
\vert GT_{if} \vert^2 {{N_i^p N_f^h}\over{(2 j_f + 1)}}
\label{strength}
\end{equation}
where $\vert GT_{if} \vert = \langle f \vert GT \vert i
\rangle$ is a single particle transition
matrix element between the states,
$N_i^p$ is the occupation number of the initial
level, $i$, and $N_f^h/(2 j_f + 1)$ is the fractional number
of holes in the final level, $f$.  Eq. (\ref{strength}) is subject to the
usual GT selection rules.  Similar formulae have been used for charged
current GT strength functions \cite{FFN}.  While detailed shell model
studies have revealed inadequacies in such an approach for
charged current interactions \cite{aufderheide} as well
as for the neutral currents \cite{ressell2}, the above
parameterization is quite useful for revealing general trends.
For the modified operator of eq. (\ref{GTop}) with $G_A^s \neq 0$,
eq. (\ref{strength}) must be altered:
\begin{eqnarray}
&&B(GT_0) \propto \vert^2 \sum_{p} \sum_{i,f} \vert GT_{if}(p) \vert^2
{{N_i^p(p) N_f^h(p)}\over{(2 j_f(p) + 1)}}
+ \nonumber \\
&& \;\;\;\;\;\;\;\;\;\;\;\;\; \sum_{n} \sum_{i,f} \vert GT_{if}(n) \vert^2
{{N_i^p(n) N_f^h(n)}\over{(2 j_f(n) + 1)}}
\label{modstrength}
\end{eqnarray}
Here, $p(n)$ denotes the proton (neutron) contribution to the
strength.  We see from eq. (\ref{modstrength}) that, since
$\vert GT_{if}(p) \vert^2 > \vert GT_{if}(n) \vert^2$, the
strength can be significantly altered for $G_A^s \neq 0$.
However, for nuclei with $N \approx Z$ the effects of the differing
matrix elements will tend to cancel.  For nuclei with
$\vert N - Z \vert \gg 0$, fairly significant effects
could be observed.  Since many nuclei in the pre-collapse and
collapse phases of a massive star's life cycle have
$N \gg Z$, weak inelastic neutral current processes could
undergo important changes.

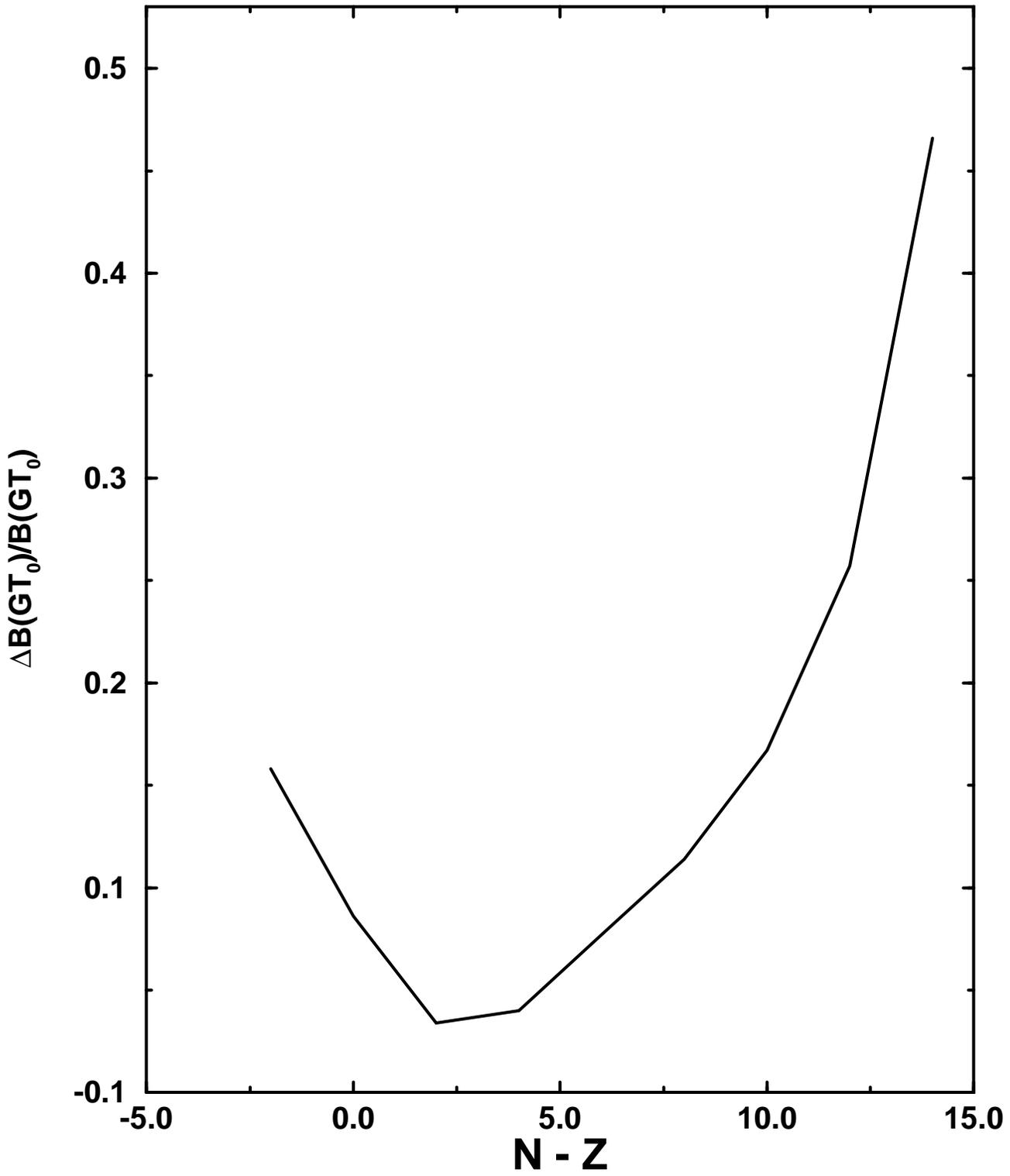
\begin{figure}
\caption[Figure 2]{The change in the Gamow-Teller strength in a
series of iron isotopes ($Z = 26$) due to
the strangeness in the nucleon as a function of $N - Z$.}
\label{dbgt0}
\end{figure}

A close examination of eq. (\ref{modstrength})
 reveals several competing effects.
As $N - Z$ increases, the naive expectation is for $B(GT_0)$ to
decrease (relative to the $G_A^s = 0$ case) because of
the decrease in $GT_{if}(n)$.  Looking at Table \ref{table2}, we see
that this trend occurs for $N - Z = -2$ to 4.  For $N - Z > 4$,
the strength, in iron, increases.  This can be traced to the
fact that the $1f_{5/2}$ and $2p_{1/2}$ neutron shells are
starting to become occupied.  This reduces the fractional
number of holes available for the transition.  This
is the well known effect of Pauli shell blocking \cite{fuller}.
This shell blocking becomes increasingly important as the neutrons
approach shell closure ($N = 40$).  For a completely
closed neutron $fp$ shell, Table \ref{table2} shows that inclusion
of strange neutral currents leads to a 42 \% increase in
the total strength.  In $^{66}$Fe, the strength is purely due to
proton transitions.  In figure \ref{dbgt0} we plot
$\Delta B(GT_0)/B(GT_0)$ against $N - Z$ for the series of
iron isotopes considered.  The competition between the altered
matrix elements and shell blocking is clearly visible.
We see that a significant change in $B(GT_0)$ can occur in nuclei
with a large neutron excess if strange quarks carry a
reasonable fraction of the nucleon's spin.

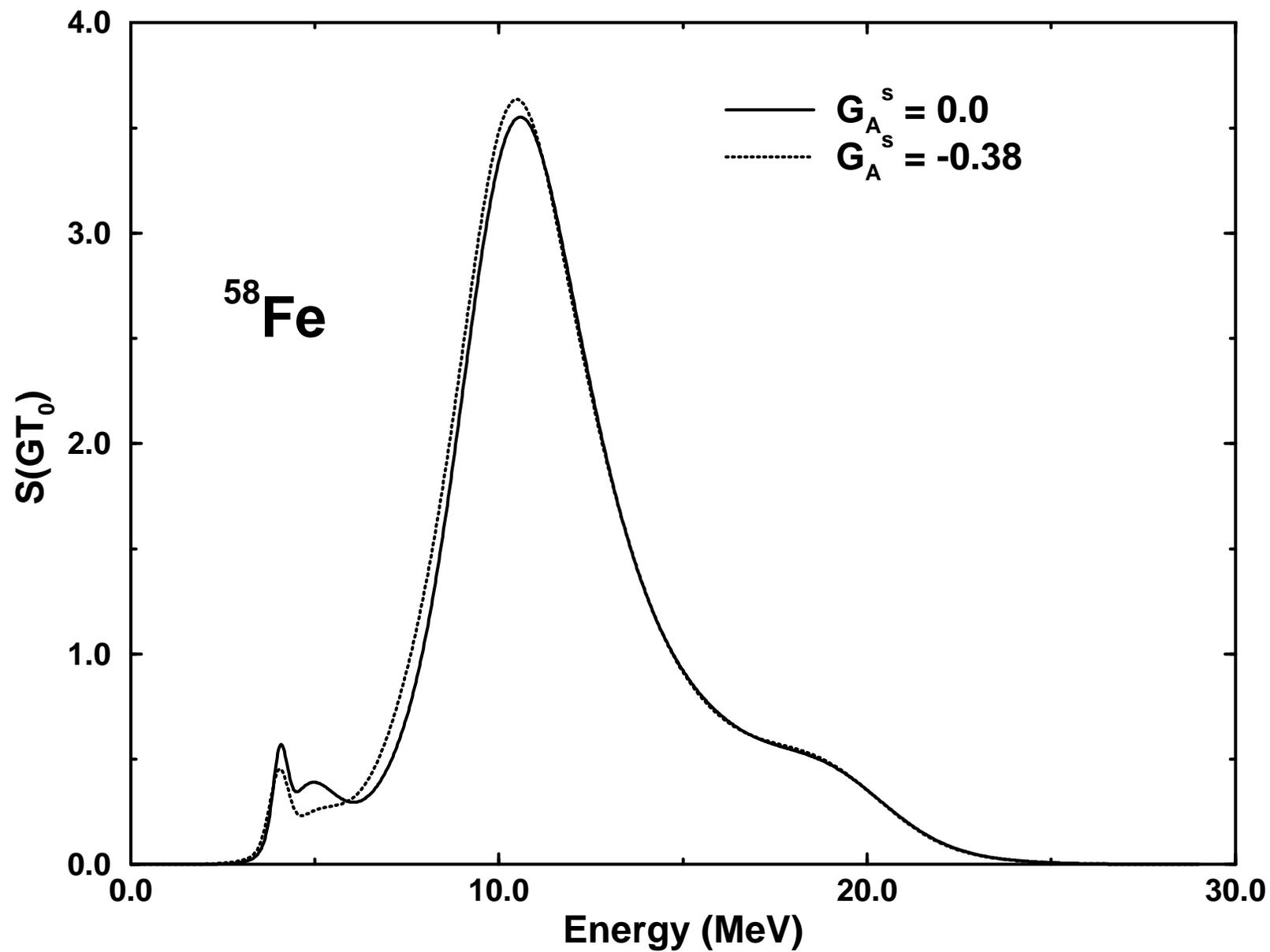
\begin{figure}
\caption[Figure 3]{The total neutral current Gamow-Teller strength
function for $^{58}$Fe.  The solid line is the result using the
standard ($G_A^s = 0$) result.  The dashed line shows the effects
of including $G_A^s = -0.38$ in the operator of eq. (\ref{GTop}).}
\label{fe58strength}
\end{figure}

In Table \ref{table3} we present the energy weighted centroid of the
neutral current strength for the nuclei considered.  No
obvious correlation of the centroid with $N - Z$ is
apparent.  Isotopes where the neutron transitions might be
expected to dominate (large $N_i^p(n)$ {\it and}
$N_f^h(n)/(2 j_f(n) + 1)$) do seem to have slightly negative
centroid shifts.  Isotopes dominated by proton
transitions (small $N_i^p(n)$ {\it or}
$N_f^h(n)/(2 j_f(n) + 1)$) tend to have
a positive shift.  In figure \ref{fe58strength} we present the
total strength function for $^{58}$Fe with and without a
contribution due to strange quarks in the nucleon.  This strength
function is derived from transitions from 30 approximate
eigenstates obtained by performing
Lanczos iterations upon the Collective Gamow-Teller state
\cite{mathews}.
Transitions which are not converged are spread out over a
Gaussian distribution with the appropriate width
obtained from the computed second moments of the eigenstates.
\cite{ressell2,mathews,bloom}.

Although we have only considered the effects of increasing $N - Z$ for a
series of iron isotopes,  there is nothing special about iron.
We therefore expect similar behavior for most of the elements present
during supernova core collapse.
One set of elements where significant changes in $B(GT_0)$
might occur are the $T = 0$ ($N = Z$) nuclei.  These nuclei
will not be abundant in the collapsing core but will be present
in the outer envelopes of the star.  As mentioned earlier,
the presence of a non-zero $G_A^s$ allows the GT
operator to mediate $T = 0 \rightarrow T = 0$ transitions.
This has the effect of re-arranging the low lying GT strength.
We now examine the magnitude of this effect.

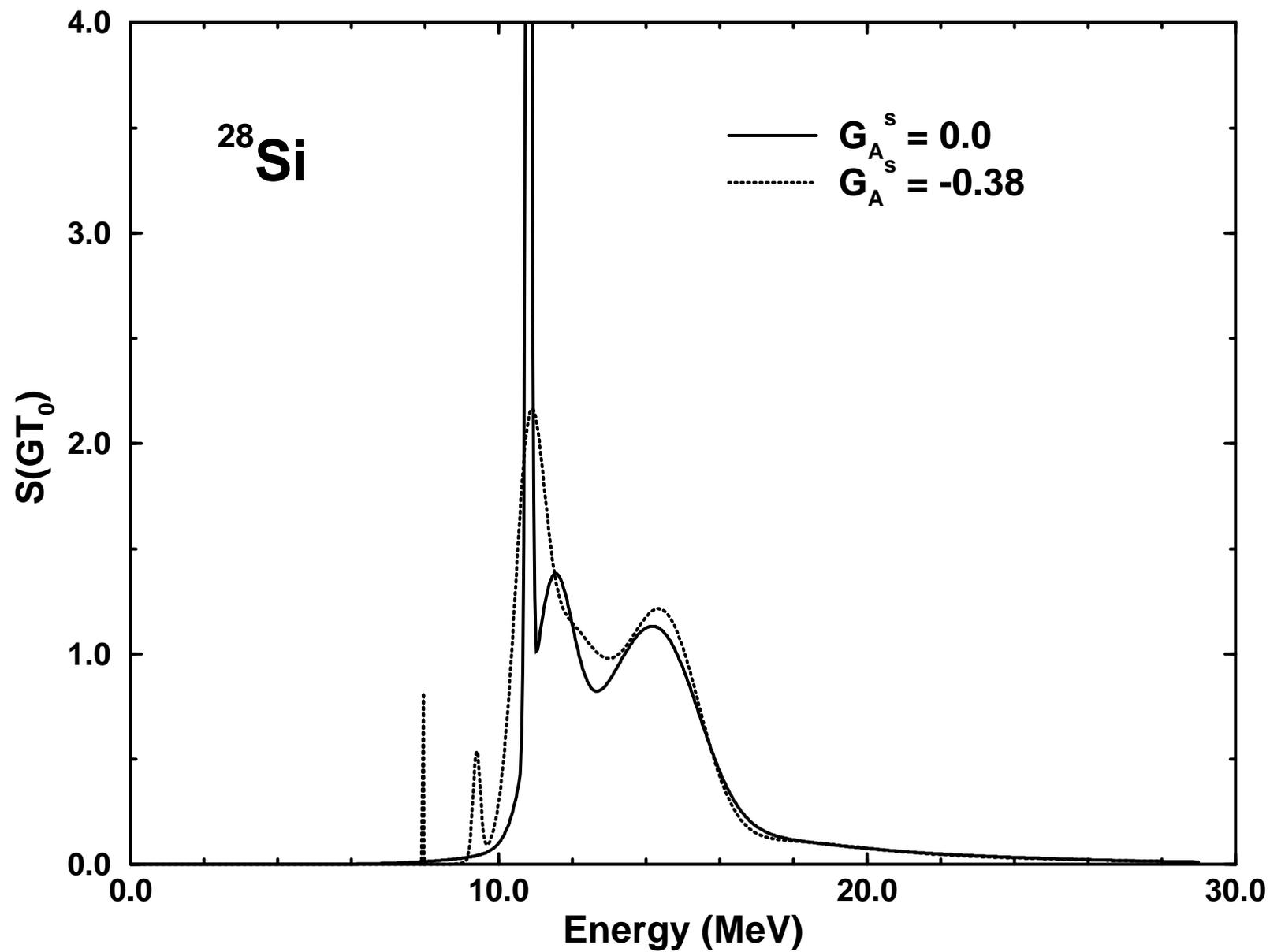
\begin{figure}
\caption[Figure 4]{The total neutral current Gamow-Teller strength
function for $^{28}$Si.  The solid line is the result using the
standard ($G_A^s = 0$) result.  The dashed line shows the effects
of including $G_A^s = -0.38$ in the operator of eq. (\ref{GTop}).  Note the
two $T = 0 \rightarrow T = 0$ transitions at 7.94 MeV and 9.40
MeV which are not present in the standard result.  Also note that
the change in appearance in the $T = 1$ peak at 10.81 MeV is primarily
due to the altered convergence properties of the Lanczos iterations
used to obtain the strength distribution.}
\label{si28strength}
\end{figure}

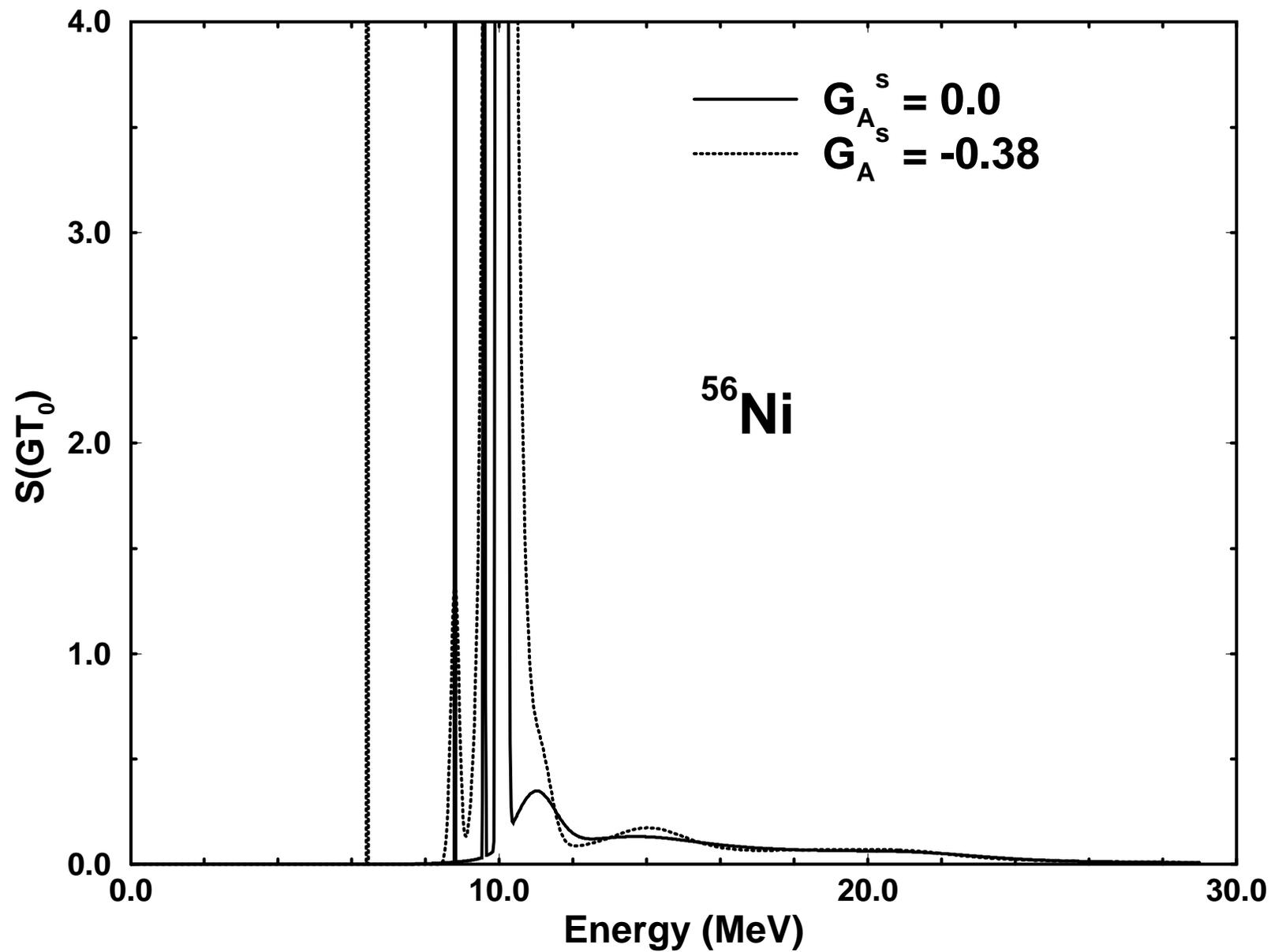
\begin{figure}
\caption[Figure 5]{The total neutral current Gamow-Teller strength
function for $^{56}$Ni.  The solid line is the result using the
standard ($G_A^s = 0$) result.  The dashed line shows the effects
of including $G_A^s = -0.38$ in the operator of eq. (\ref{GTop}).  Note the
$T = 0 \rightarrow T = 0$ transition at 6.43 MeV. Also note that
the change in appearance of the $T = 1$ peaks near 10 MeV is primarily
due to the altered convergence properties of the Lanczos iterations
used to obtain the strength distribution.}
\label{ni56strength}
\end{figure}

In figures \ref{si28strength} and \ref{ni56strength}
we present the strength distributions for
$^{28}$Si and $^{56}$Ni,
both $T = 0$ nuclei which will be abundant as a residue of thermonuclear
burning in shells within and just above
the collapsing core of a massive star.  The solid line in each figure
is the standard $G_A^s = 0$ strength and the dotted line is the
strength with $G_A^s = -0.38$.  In each figure at least one new
low lying isoscalar transition is apparent.  There is also a shift in the
shape of the resonance.  This shift is primarily due to the
change in convergence of the Lanczos vectors used to
determine the strength distribution.  For a more detailed
discussion of the procedure used to obtain $B(GT_0)$ see refs.
\cite{ressell2,mathews,bloom}.

Tables \ref{table2} and \ref{table3}
show the quantitative shifts in the total
strength and centroid of the $T = 0$ nuclei respectively.
There is a small, relatively constant, increase in the total
strength of about 3 \%.  The one exception, $^{14}$N, has
roughly a 6 \% increase and is the only odd-odd nucleus
considered here.  We also see a fairly uniform slight decrease
in the strength centroid for most of the nuclei considered.
Thus,
we see, that the redistribution of GT strength in
$T = 0$ nuclei is not likely to be a large effect.

To confirm the above statement, we must look at a real physical
process.  We choose the process $A^* \rightarrow
A \nu {\overline{\nu}}$, neutral current de-excitation of a
nucleus, in a hot stellar environment.  This process is highly
sensitive to the distribution of GT strength and hence should be
an excellent indicator of the possible importance of
$G_A^s \neq 0$ upon inelastic neutral current scattering processes in
supernovae.  This process has recently been considered in detail
in ref. \cite{ressell2}. Here we sketch the details of the
calculation.

We consider only decays to the ground state of the nucleus.  Refs.
\cite{ressell2,bh} show that these decays dominate
the rate for temperatures, $T$, less than about $1.5$ MeV for
$fp$ shell nuclei.
A more complete set of final states is required at higher $T$, but
the effect we wish to emphasize is well illustrated by
decay to this single state.

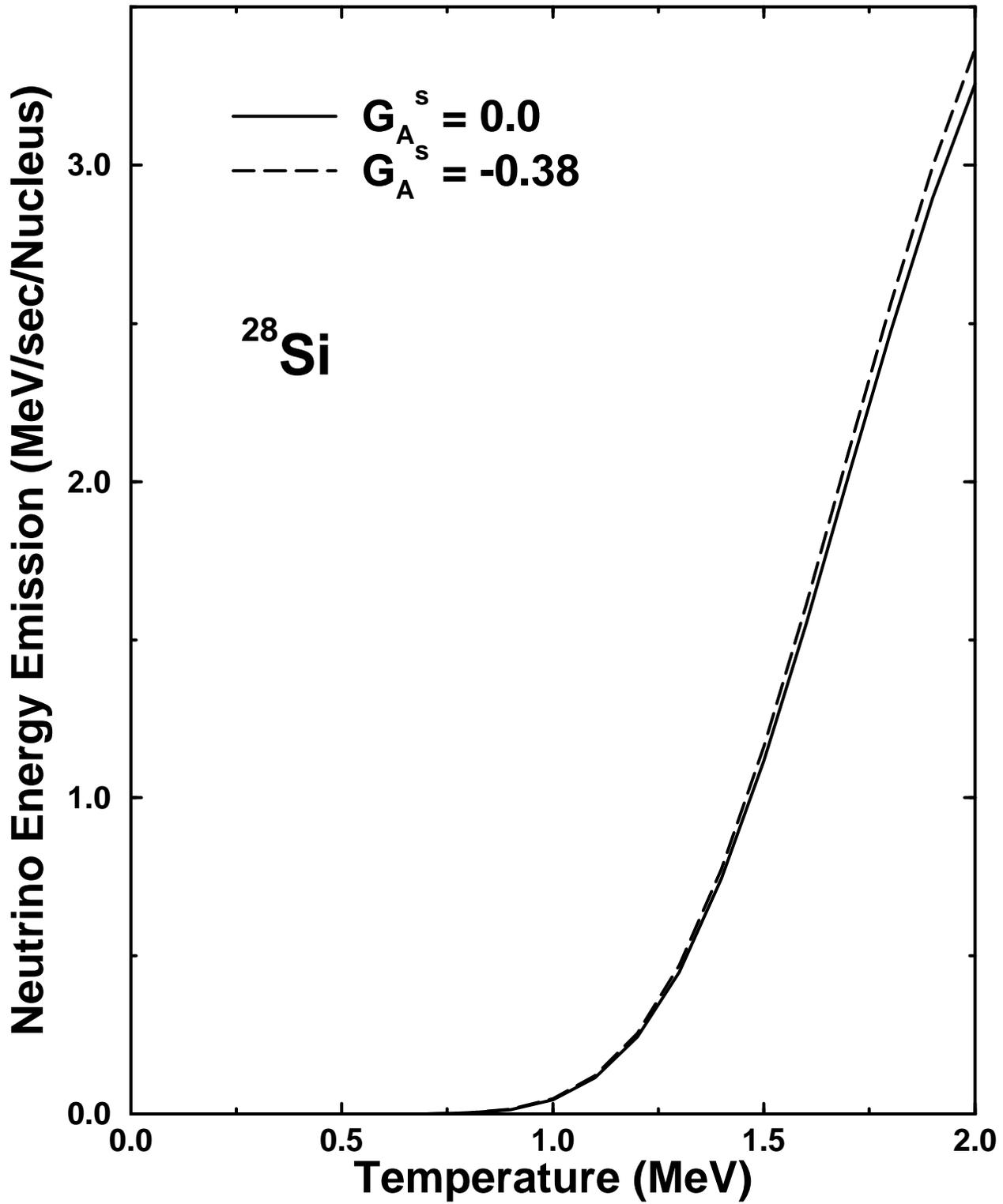
\begin{figure}
\caption[Figure 6]{The neutrino energy emission rate
for the process $A^* \rightarrow A \nu{\overline\nu}$
(${\dot\epsilon}_{\nu{\overline\nu}}$, eq. (\ref{epsdot})) as a function
of Temperature for $^{28}$Si.  The rate includes only decay to
the ground state.}
\label{si28rate}
\end{figure}

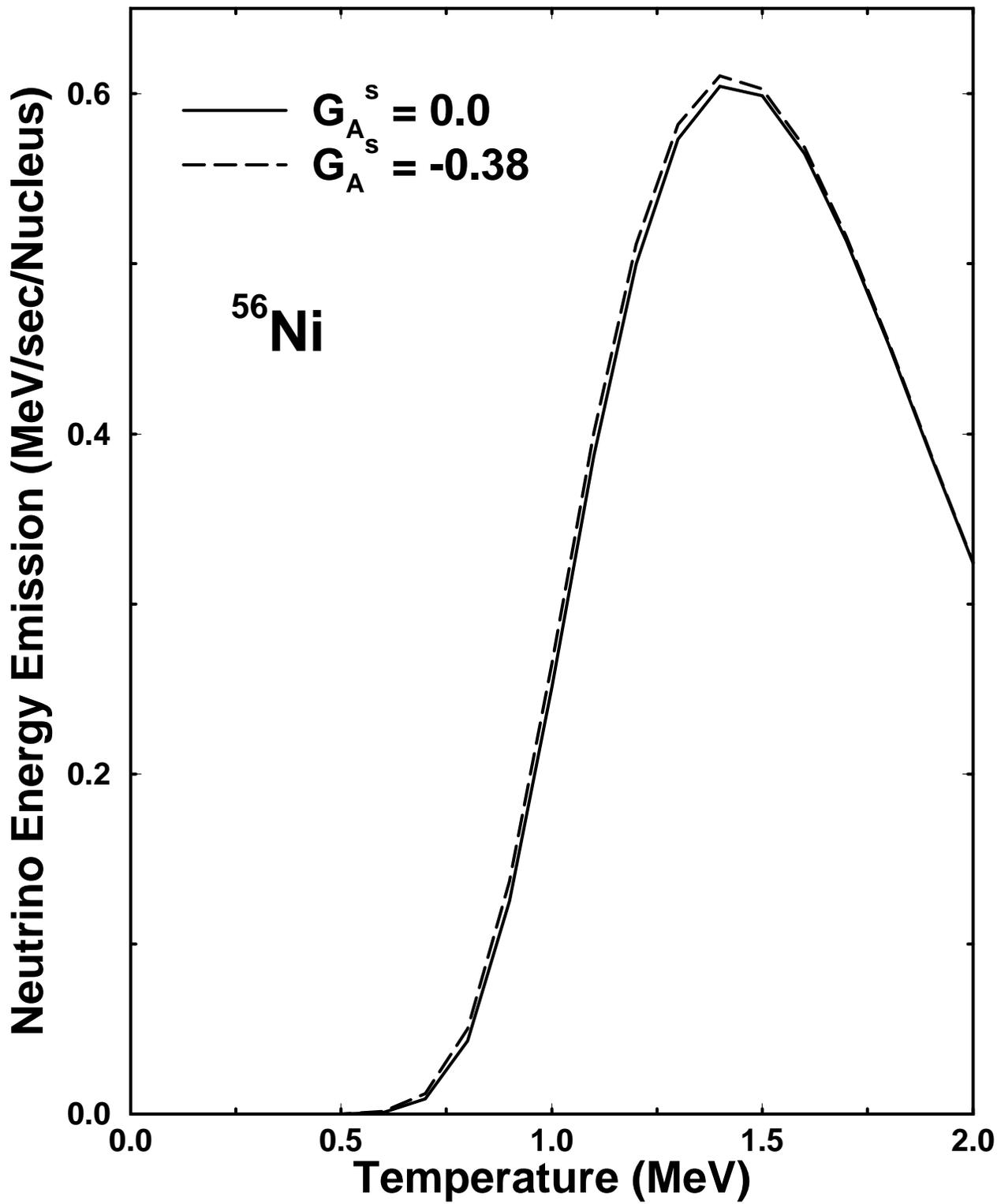
\begin{figure}
\caption[Figure 7]{The neutrino energy emission rate
for the process $A^* \rightarrow A \nu{\overline\nu}$
(${\dot\epsilon}_{\nu{\overline\nu}}$, eq. (\ref{epsdot})) as a function
of Temperature for $^{56}$Ni.  The rate includes only decay to
the ground state.}
\label{ni56rate}
\end{figure}

The neutrino pair energy emission rate from ground-state
transitions of a thermal population
of nuclear states is
\begin{eqnarray}
&&{\dot{\epsilon}}_{\nu {\overline{\nu}}} = 3.33 \times 10^{-4}
{g_a^2\over{4}} \sum_i \vert GT_{if} \vert^2 (E_i - E_f)^6
\nonumber \\
&& \;\;\;\;\;\;\;\;\; \times {(2 J_i + 1) e^{-E_i/T}\over{G(Z,A,T)}}
{\rm{MeV sec}^{-1}} {\rm{nucleus}}^{-1}.
\label{epsdot}
\end{eqnarray}
Here, $GT_{if}$ is now the matrix element connecting the shell model
states ($i =$ initial, $f =$ final = ground state), $E_i$ is the
excited state energy, $E_f = 0$ is the final state's energy,
$J_i$ is the initial state angular momentum, and $G(Z,A,T)$
is the nuclear partition function.  We see that there are two effects
of a non-zero $G_A^s$. First, from the
magnitude of the strength as represented by $\vert GT_{if}
\vert^2$. Second, from the {\it location} of the strength through
the factor $(E_i - E_f)^6$.  This latter effect could be
particularly important for $T = 0$ nuclei where new states
become available.  In Table \ref{table4} we present results for
${\dot{\epsilon}}_{\nu {\overline{\nu}}}$, for several
of the calculated nuclei at $T = 1$ MeV.  In figures \ref{si28rate}
and \ref{ni56rate}
we show ${\dot{\epsilon}}_{\nu {\overline{\nu}}}$ as a
function of temperature for $^{56}$Ni and $^{28}$Si.
The turn over at $T \sim 1.5$ MeV in $^{56}$Ni is an artifact
of only considering ground state decay.  It indicates  the need to consider
decay to more states than just the ground state
above this temperature.
Looking at Table \ref{table4} and figures \ref{si28rate} and
\ref{ni56rate}, it is once again
evident that the effect of the strangeness content of the nucleon
is typically of order a few percent for most nuclei.

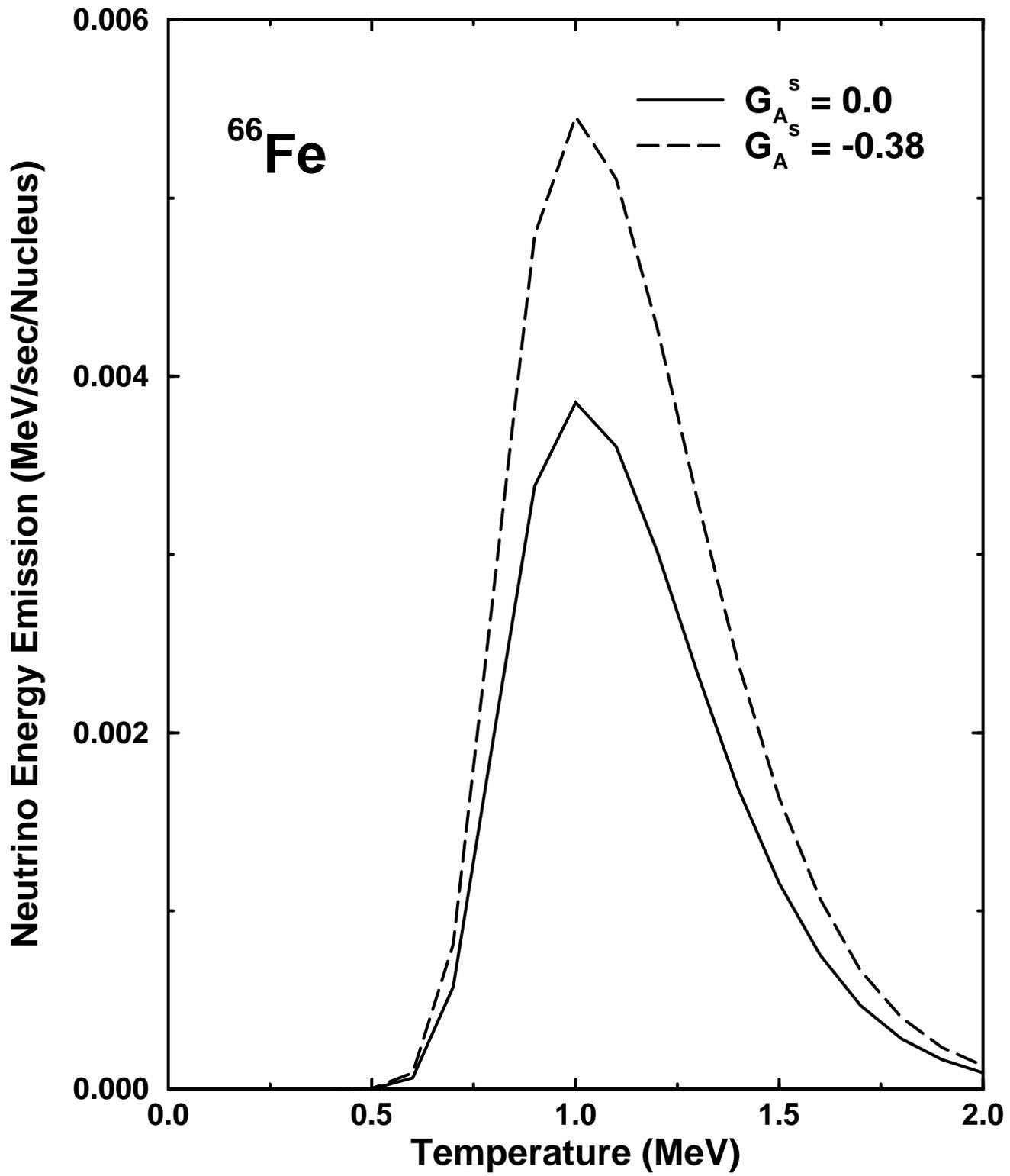
\begin{figure}
\caption[Figure 8]{The neutrino energy emission rate
for the process $A^* \rightarrow A \nu{\overline\nu}$
(${\dot\epsilon}_{\nu{\overline\nu}}$, eq. (\ref{epsdot})) as a function
of Temperature for $^{66}$Fe.  The rate includes only decay to
the ground state.  The large enhancement over the $G_A^s = 0$
case can be traced to the fact that this nucleus has no allowed
neutron transitions in this model space.}
\label{fe66rate}
\end{figure}

$^{66}$Fe is somewhat of an exception to the above statement.
For this nucleus $N = 40$ and there are
{\it no} allowed neutron transitions in our model space.
Thus, we see the full effect of the enhanced $\nu p$ interaction
strength.  This is already obvious from the large value of
$\Delta B(GT_0)/B(GT_0)$ in Table \ref{table2}.  This produces
a commensurate increase in
${\dot{\epsilon}}_{\nu {\overline{\nu}}}$.
Table \ref{table4} shows that
${\dot{\epsilon}}_{\nu {\overline{\nu}}}$ increases by
$\sim 40 \%$ relative to the $G_A^s = 0$ value at
$T \sim 1$ MeV.  This is comparable to the naive
estimates made at the beginning of section III.  In figure
\ref{fe66rate} we show ${\dot{\epsilon}}_{\nu {\overline{\nu}}}$ for the
decay to the ground state as a function of temperature.  Unfortunately,
this large enhancement of neutrino energy emission is likely
to occur only for nuclei with $N = 40$.  If at any point in the
collapse this condition is encountered, cooling due to neutral
current de-excitation could be greatly enhanced.  Additionally,
$fp$ shell nuclei with $N = 40$ have no allowed electron capture
strength, further increasing the significance of this process
in this regime.  Of course, excitations out of our model space
into the $sdg$ shell would tend to smooth over this enhancement
by allowing the neutrons to once again contribute to the rate.

Other inelastic neutral current processes, such as both up and
down $\nu A$ scattering and $\nu$ pair annihilation onto a nucleus
($\nu{\overline{\nu}}A \rightarrow A^*$) will respond
similarly to a non-zero $G_A^s$.  For these processes, there
will be identical shifts due to the altered matrix elements,
$GT_{if}$, but the phase space will scale less steeply than the
$(E_i - E_f)^6$ factor encountered in eq. (\ref{epsdot}).
Thus, we see that the
strangeness content of the nucleon plays, at most, only a minor
role in the energy exchange and transport in the
collapsing cores of massive stars.  (If the star passes through
a regime where $N = 40$ nuclei are extremely abundant, the strangeness
might play a very important role.) The equally interesting
question of the effect of $G_A^s \neq 0$ upon
$\nu$-process nucleosynthesis has been previously investigated in
ref. \cite{kolbe}.

We close this section by briefly mentioning the possibility of
using inelastic neutral current $\nu A$ scattering to determine
$G_A^s$.  This possibility has been explored for several nuclei
in refs. \cite{sky,hkpw}. One idea is to
use a $T = 0$ nucleus and look for a neutrino mediated transition
to a $T = 0$ excited state via its subsequent decay(s) to the
ground state.  In the regime where the allowed approximation
applies, the $\nu$ excitation cross section is proportional to
$\vert G_A^s \vert^2$.  Hence,
an excitation to a $T = 0$ state would determine a
value for $G_A^s$ (e.g. provided a sufficiently detailed and accurate
nuclear model is available).
The advantage
of this method lies in the fact that the measurement is made in
the $Q^2 = 0$ limit where one would be
measuring $\Delta s$ in the regime of interest.  This stands
in contrast to the accelerator experiments (e.g. EMC \cite{emc},
SMC \cite{smc}]) which measure $\Delta s$ at large $Q^2$.  The
results then have to be extrapolated to the $Q^2 = 0$ case.
Since this extrapolation is through the non-perturbative regime,
a great deal of uncertainty is introduced.  While this method is
extremely attractive from a theoretical standpoint, it
is a challenging experimental task at best.

\section{CONCLUSIONS} \label{sec:conclusions}

In this paper we have examined the consequences of a non-negligible
value of $\Delta s$ (or equivalently, $G_A^s$) upon
inelastic $\nu A$ interactions.  We have focused upon
neutral current Gamow-Teller processes for a number of nuclei
that might be relevant for core collapse supernovae.  All
of the calculations have been done in the nuclear shell
model in order to obtain an accurate representation of the
Gamow-Teller strength, $B(GT_0)$.  We have focused upon two aspects
of the modified GT operator, eq. (\ref{GTop}).  The first of these is the
deviation, from the $G_A^s = 0$ case, as the value of $N - Z$
changes.  The second is the changes in $B(GT_0)$ for $T = 0$
nuclei which result from the isoscalar piece in eq. (\ref{GTop}).  Most
of our discussion focuses upon the strength function, $B(GT_0)$,
but we have also examined a real physical process which may be
important in core collapse supernovae, i.e. the process
$A^* \rightarrow A\nu{\overline \nu}$.  This nuclear de-excitation
should be especially sensitive to $G_A^s$ due to its steep
dependence upon the excitation energy.

In the initial collapse of a massive star's core, nuclei will
become very neutron rich, $N \gg Z$.  The inclusion of a
non-zero $G_A^s$ would then change the ratio of $\nu p$ and
$\nu n$ interaction strengths leading to a redistribution in
GT strength.  This redistribution should become more pronounced
as $N - Z$ increases.  In tables II and III we presented our
results for both $B(GT_0)$ and its centroid with $G_A^s = 0$
and $G_A^s = -0.38$ for a series of iron isotopes with
$-2 \le N - Z \le 14$.  We found that $B(GT_0)$ initially
decreased as $N$ increased, in line with expectations.  As
the neutron $1f_{5/2}$ and $2p_{1/2}$ orbitals began to fill, shell
blocking became important.  At this point, proton transitions
began to dominate the strength.  Thus for large $N - Z$ there
was a steady increase in $B(GT_0)$.  This change was typically
of order 10 \% but reached $\sim 40 \%$ for $^{66}$Fe.
Since there were no new transitions added, just a reweighting
of those already present, there was very little change in the
position of the strength's centroid. This implies that the change
in $\dot\epsilon_{\nu{\overline \nu}}$, eq. (\ref{epsdot})
is predominately controlled
by the change in $B(GT_0)$ and the large phase space factor does
not come into play.

For $T = 0$ nuclei, the story is somewhat different.  For
these nuclei $G_A^s \neq 0$, which leads to a violation of the
standard no $T = 0 \rightarrow T = 0$ selection rule for
GT transitions.  Thus, new interaction channels become
available and a significant rearrangement of the strength becomes
possible.  We studied $B(GT_0)$ in a number of $T = 0$ nuclei
and found only about a 3--6 \% change in
$B(GT_0)$ and only a small shift for its centroid.  The
results are presented in Tables \ref{table2} and \ref{table3}.  We find that
despite new channels opening up, the total strength is only
slightly modified.  Nevertheless, these new channels do, perhaps, present
an intriguing method for measuring $\Delta s$
even though the strength
present in these new channels is insufficient to cause a major
change in the energy emission rate for the nuclear de-excitation
process considered.

We close by noting that a non-zero $G_A^s$ seems unlikely to
produce a significant change in calculated neutral current
GT processes \cite{ressell2,bh}.  The effects
are not entirely negligible but will not severely change the
dominant neutrino energy emission mechanisms.  This is especially
true in that we have used a value of $\vert G_A^s \vert$ which
is probably too large (the SMC finds a smaller, but non-zero,
value \cite{smc}). Also, we have quenched $g_A$ but not
$G_A^s$.  In that sense, the results presented here should be
regarded as upper limits.

\acknowledgements

Work at LLNL was performed under the auspices of the U.S. Department
of Energy under contract W-7405-ENG-48
M.A. acknowledges support from the U.S. Department of Energy
under contract DOE-AC02-76-ERO-3071.  M.T.R. acknowledges support from
the Weingart foundation and the U.S. National Science Foundation
under Grants PHY94-12818 and PHY94-20470.
Work at the University of Notre Dame supported by DOE Nuclear
Theory Grant DE-FG02-95ER-40934.

\newpage
\onecolumn
\squeezetable
\widetext
\begin{table}
\caption[Table I]{The model spaces used for calculating $B(GT_0)$ for
$fp$ shell nuclei.  Columns 2 and 3 list the parent model space and
m-scheme dimension.  Columns 4 and 5 list the same for the daughter
nucleus' space.}
\begin{tabular}{ccccc}
Nucleus & Parent Model Space & Dim. & Daughter Model Space
& Dim. \\
 & & & & \\
\tableline
 & & & & \\
$^{50}$Fe & $(1f_{7/2})^{10,9}(2p_{3/2}2p_{1/2}1f_{5/2})^{0,1}$ & 5350 &
$(1f_{7/2})^{10,9,8}(2p_{3/2}2p_{1/2}1f_{5/2})^{0,1,2}$ & 67948 \\
$^{52}$Fe & $(1f_{7/2})^{12,11}(2p_{3/2}2p_{1/2}1f_{5/2})^{0,1}$ & 3160 &
$(1f_{7/2})^{12,11,10}(2p_{3/2}2p_{1/2}1f_{5/2})^{0,1,2}$ & 57710 \\
$^{54}$Fe & $(1f_{7/2})^{14,13}(2p_{3/2}2p_{1/2}1f_{5/2})^{0,1}$ & 328 &
$(1f_{7/2})^{14,13,12}(2p_{3/2}2p_{1/2}1f_{5/2})^{0,1,2}$ & 10620 \\
$^{56}$Fe & $(1f_{7/2})^{14}(2p_{3/2}2p_{1/2}1f_{5/2})^{2}$ & 200 &
$(1f_{7/2})^{14,13}(2p_{3/2}2p_{1/2}1f_{5/2})^{2,3}$ & 8738 \\
$^{58}$Fe & $(1f_{7/2})^{14}(2p_{3/2}2p_{1/2}1f_{5/2})^{4}$ & 1392 &
$(1f_{7/2})^{14,13}(2p_{3/2}2p_{1/2}1f_{5/2})^{4,5}$ & 46310 \\
$^{60}$Fe & $(1f_{7/2})^{14}(2p_{3/2}2p_{1/2}1f_{5/2})^{6}$ & 2542 &
$(1f_{7/2})^{14,13}(2p_{3/2}2p_{1/2}1f_{5/2})^{6,7}$ & 72298 \\
$^{62}$Fe & $(1f_{7/2})^{14}(2p_{3/2}2p_{1/2}1f_{5/2})^{8}$ & 1392 &
$(1f_{7/2})^{14,13}(2p_{3/2}2p_{1/2}1f_{5/2})^{8,9}$ & 35482 \\
$^{64}$Fe & $(1f_{7/2})^{14,13}(2p_{3/2}2p_{1/2}1f_{5/2})^{10,11}$ & 4638 &
$(1f_{7/2})^{14,13,12}(2p_{3/2}2p_{1/2}1f_{5/2})^{10,11,12}$ & 37360 \\
$^{66}$Fe & $(1f_{7/2})^{14,13,12,11}(2p_{3/2}2p_{1/2}1f_{5/2})^{12,13,14,15}$
 & 1710 &
$(1f_{7/2})^{14,13,12,11,10}(2p_{3/2}2p_{1/2}1f_{5/2})^{12,13,14,15,16}$
& 3102 \\
$^{56}$Ni & $(1f_{7/2})^{16,15,14}(2p_{3/2}2p_{1/2}1f_{5/2})^{0,1,2}$ & 1353 &
$(1f_{7/2})^{16,15,14,13}(2p_{3/2}2p_{1/2}1f_{5/2})^{0,1,2,3}$ & 34593 \\
\end{tabular}
\label{table1}
\end{table}

\mediumtext
\begin{table}
\caption[Table II]{The change in the total neutral current GT
strength, $B(GT_0)$.  Column 3 lists $B(GT_0)$ with $G_A^s = 0$,
column 4 lists $B(GT_0)$ with $G_A^s = -0.38$, and column 5 lists
the change between the two divided by the $G_A^s = 0$ value.}
\begin{tabular}{ccccc}
Nucleus & $N - Z$ & $B(GT_0)\vert_{G_A^s = 0}$ & $B(GT_0)\vert_{G_A^s =
-0.38}$ & $\Delta B(GT_0)$ / $B(GT_0)\vert_{G_A^s = 0}$ \\
 & & & & \\
\tableline
 & & & & \\
$^{50}$Fe & -2 & 17.3521 & 19.2245 & 0.108 \\
$^{52}$Fe & 0 & 20.3881 & 21.1276 & 0.036 \\
$^{54}$Fe & 2 & 24.0775 & 23.6927 & -0.016 \\
$^{56}$Fe & 4 & 23.7421 & 23.5076 & -0.01 \\
$^{58}$Fe & 6 & 21.4906 & 22.0658 & 0.027 \\
$^{60}$Fe & 8 & 19.6358 & 20.8826 & 0.064 \\
$^{62}$Fe & 10 & 17.1889 & 19.2029 & 0.117 \\
$^{64}$Fe & 12 & 14.3119 & 17.2748 & 0.207 \\
$^{66}$Fe & 14 & 8.4445 & 11.9583 & 0.416 \\
 & & & & \\
\tableline
 & & & & \\
$^{56}$Ni & 0 & 23.9834 & 24.7462 & 0.032 \\
$^{28}$Si & 0 & 7.7844 & 8.0314 & 0.032 \\
$^{24}$Mg & 0 & 4.6568 & 4.8198 & 0.035 \\
$^{20}$Ne & 0 & 10.9113 & 11.3027 & 0.036 \\
$^{14}$N & 0 & 4.6276 & 4.8938 & 0.058 \\
$^{12}$C & 0 & 2.0906 & 2.1555 & 0.031 \\
\end{tabular}
\label{table2}
\end{table}

\begin{table}
\caption[Table III]{The change in the location (in MeV) of the
energy weighted centroid of the GT strength for all of the nuclei
studied.  Column 3 lists the centroid with $G_A^s = 0$, column
4 lists the centroid with $G_A^s = -0.38$, and column 5 lists the
change in the position.}
\begin{tabular}{ccccc}
 Nucleus & $N - Z$ & Centroid($G_A^s$ = 0) & Centroid($G_A^s$ = -0.38)
& $\Delta$Centroid\\
 & & & & \\
\tableline
 & & & & \\
$^{50}$Fe & -2 & 12.4738 & 12.5194 & 0.041 \\
$^{52}$Fe & 0 & 13.0520 & 12.9440 & -0.108 \\
$^{54}$Fe & 2 & 13.2015 & 13.0450 & -0.157 \\
$^{56}$Fe & 4 & 11.473 & 11.336 & -0.137 \\
$^{58}$Fe & 6 & 11.7908 & 11.7161 & -0.075 \\
$^{60}$Fe & 8 & 11.7007 & 11.7458 & 0.045 \\
$^{62}$Fe & 10 & 11.2949 & 11.4172 & 0.122 \\
$^{64}$Fe & 12 & 11.6246 & 11.6335 & 0.009 \\
$^{66}$Fe & 14 & 11.1195 & 11.1195 & 0.0 \\
 & & & & \\
\tableline
 & & & & \\
$^{56}$Ni & 0 & 10.3315 & 10.2217 & -0.110 \\
$^{28}$Si & 0 & 13.5540 & 13.4686 & -0.075 \\
$^{24}$Mg & 0 & 13.4006 & 13.3583 & -0.042 \\
$^{20}$Ne & 0 & 15.8091 & 15.8346 & 0.026 \\
$^{14}$N & 0 & 9.8878 & 9.6560 & -0.232 \\
$^{12}$C & 0 & 15.6493 & 15.5642 & -0.085 \\
\end{tabular}
\label{table3}
\end{table}

\begin{table}
\caption[Table IV]{The value of the energy emission rate for the
process $A^* \rightarrow A \nu{\overline{\nu}}$,
${\dot{\epsilon}}_{\nu {\overline{\nu}}}$ in eq. \ref{epsdot}, in
MeV/sec/Nucleus for several nuclei.}
\begin{tabular}{ccccc}
 & & & & \\
 Nucleus & & $G_A^s = 0$ & $G_A^s = -0.38$ & Change \\
 & & & & \\
\tableline
 & & & & \\
$^{56}$Fe & & 0.03944 & 0.04168 & 5.7 \% \\
$^{58}$Fe & & 0.02228 & 0.02362 & 6.0 \% \\
$^{62}$Fe & & 0.01093 & 0.01209 & 10.1 \% \\
$^{64}$Fe & & 0.00627 & 0.00744 & 18.6 \% \\
$^{66}$Fe & & 0.00385 & 0.00546 & 41.6 \% \\
$^{56}$Ni & & 0.25095 & 0.26501 & 5.6 \% \\
$^{28}$Si & & 0.04577 & 0.04882 & 6.7 \% \\
\end{tabular}
\label{table4}
\end{table}

\end{document}